\documentclass[12pt,preprint]{aastex}  

\shorttitle{A Hot and Massive Disk around IRAS 20126+4104} %
\shortauthors{Chen et al.} %

\begin{document} %
\title{A Hot and Massive Accretion Disk around the High-Mass Protostar IRAS~20126+4104} %

\author{Huei-Ru Vivien Chen\altaffilmark{1,2}, Eric Keto\altaffilmark{3}, Qizhou Zhang\altaffilmark{3}, T.~K. Sridharan\altaffilmark{3}, Sheng-Yuan Liu\altaffilmark{2}, and Yu-Nung Su\altaffilmark{2}} %

\altaffiltext{1}{Institute of Astronomy and Department of Physics, National Tsing Hua University, 101, Sec. 2, Kuang Fu Road, Hsinchu 30013, Taiwan; hchen@phys.nthu.edu.tw.} %
\altaffiltext{2}{Academia Sinica Institute of Astronomy and Astrophysics, P.O. Box 23-141, Taipei, 10617, Taiwan} %
\altaffiltext{3}{Harvard-Smithsonian Center for Astrophysics, 60 Garden Street, Cambridge, MA 02318, USA} %

\begin{abstract} %
We present new spectral line observations of the $\mathrm{CH_3CN}$ molecule in the accretion disk around the massive protostar IRAS 20126+4104 with the Submillimeter Array that for the first time measure the disk density, temperature, and rotational velocity with sufficient resolution ($0\farcs37$, equivalent to $\sim 600 \; \mathrm{AU}$) to assess the gravitational stability of the disk through the Toomre-$Q$ parameter.    
Our observations resolve the central $2000 \; \mathrm{AU}$ region that shows steeper velocity gradients with increasing upper state energy, indicating an increase in the rotational velocity of the hotter gas nearer the star.  
Such spin-up motions are characteristics of an accretion flow in a rotationally supported disk.  
We compare the observed data with synthetic image cubes produced by three-dimensional radiative transfer models describing a thin flared disk in Keplerian motion enveloped within the centrifugal radius of an angular-momentum-conserving accretion flow.   
Given a luminosity of $1.3 \times 10^4 \; L_\odot$, the optimized model gives a disk mass of $1.5 \; M_\odot$ and a radius of $858 \; \mathrm{AU}$ rotating about a $12.0 \; M_\odot$ protostar with a disk mass accretion rate of $3.9 \times 10^{-5} \; M_\odot \, \mathrm{yr^{-1}}$. 
Our study finds that, in contrast to some theoretical expectations, the disk is hot and stable to fragmentation with $Q > 2.8$ at all radii which permits a smooth accretion flow.     
These results put forward the first constraints on gravitational instabilities in massive protostellar disks, which are closely connected to the formation of companion stars and planetary systems by fragmentation.   
\end{abstract} %



\section{Introduction} %
What role accretion disks play in the formation of high-mass stars ($M \gtrsim 8 M_\odot$) remains a long-standing question.  
Circumstellar disks form naturally in the centers of rotating inflows and are a key element in the standard paradigm of the formation of Sun-like stars, providing for the growth of planetary systems.  
However, it remains debatable whether high-mass stars form in a similar fashion.  
The accretion rates in massive star formation may be high enough to induce gravitational instabilities and put disk-mediated accretion in doubt. 
Previous observations reporting disk-like accretion flows around high-mass protostars have not yet assessed the stability of the candidate disks \citep{Chini:2004bh,Patel:2005hm,JimenezSerra:2007ga,Cesaroni:2014iq,Johnston:2015jq}. 
On theoretical grounds, a very high accretion rate ($\gtrsim 10^{-4} \; M_\odot \, \mathrm{yr^{-1}}$) is required to form stars more massive than $8 \; M_\odot$.  
Stars this massive undergo rapid enough Kelvin-Helmholtz contraction that they begin hydrogen burning while still accreting \citep{Palla:1993ib}.  
A continuous resupply of fresh hydrogen is required to allow a growing massive protostar to reach the mass of a B or O star before exhausting its hydrogen fuel and leaving the main sequence \citep{Keto:2003kk}.  
The gas densities required for such rapid accretion could induce gravitational instabilities in the accretion flow that would make the disk prone to fragmentation, perhaps producing companion objects \citep{Kratter:2006jt}.  
On the other hand, the stability of a disk is determined not only by the gas density, but also the differential shear and the gas temperature which are much affected by the disk environment \citep{Durisen:2007wg}.  
High stellar luminosity ($L > 10^4 \; L_\odot$) and shock heating may keep the disk warm enough to allow accretion to proceed steadily \citep{Pickett:2000ec,Kratter:2010gs}.  
Massive disks are also subject to dynamical heating from spiral shocks, and vertical shear at the interface with the envelope \citep{Pickett:2000iu,Harsono:2011ef}.  
The stability of the disk may be assessed by the Toomre-$Q$ parameter, 
\begin{equation} %
Q = \frac{c_s \Omega}{\pi G \Sigma},    \label{eq_Q}
\end{equation} %
which compares the stabilizing effects of the temperature (sound speed $c_s$) and shear (angular velocity $\Omega$) against the clumping tendency toward instability induced by the surface density ($\Sigma$).  
Values of $Q > 1$ imply a stable disk.  
Up until now, there has been insufficient observational evidence to answer the question.  
We present new molecular line observations of the massive protostar IRAS~20126+4104 that directly address the question of the gravitational stability of massive protostellar disks.  

IRAS~20126+4104 (hereafter I20126) is a nearby \citep[$1.64 \; \mathrm{kpc}$;][]{Moscadelli:2011bo}, luminous \citep[$\sim 1.3 \times 10^4 \; L_\odot$;][]{Johnston:2011cc} high-mass protostar with a well collimated bipolar molecular outflow \citep{Zhang:1999dc,Moscadelli:2005eo,Su:2007ey,Hofner:2007do} and an accretion disk \citep{Zhang:1998fb,Cesaroni:2005eu,Cesaroni:2014iq,Xu:2012eo}.  
Previous radio and infrared observations suggest a disk mass of several $M_\odot$ and a protostellar mass of $7$ to $12 \; M_\odot$ \citep{Sridharan:2005cd,Cesaroni:2005eu,Keto:2010hj,Johnston:2011cc}.  
With such a high ratio of the disk to stellar mass, the self-gravity of the disk is significant in the accretion dynamics, and the stability of the disk is questionable.  
A previous study compared observed infrared emission against that predicted by a model of a disk-mediated accretion flow \citep{Johnston:2011cc} and found the Toomre-$Q$ high enough to allow self-consistency with their original assumption of a disk.  
Our new observations directly measure the temperature, angular velocity, and surface density to enable a direct determination of the Toomre-$Q$ and the disk stability. 

Our observations of the $\mathrm{CH_3CN}$ line emissions around I20126 with the Submillimeter Array\footnotemark[4] \citep[SMA;][]{Ho:2004bz} achieved the highest possible angular resolution ($0\farcs37$, equivalent to $\sim 600 \; \mathrm{AU}$) sufficient to spatially resolve the accretion flow.
The $\mathrm{CH_3CN}$ molecule is ideal for the identification of the hot, dense gas in the flow near the star.  
It requires high gas densities, $\gtrsim 10^4 \; \mathrm{cm^{-3}}$, for collisional excitation, and high temperatures, $\gtrsim 100 \; \mathrm{K}$, to produce detectable emission \citep{Araya:2005gb,Chen:2006iu}.  
The Doppler shifting of the molecular lines measures the rotational velocities and shear within the accretion disk.  
The emission includes multiple line transitions with different excitation energies allowing a measurement of the gas density from the line brightness and a measurement of the gas temperature from the brightness ratios.   

The present study of the high-mass protostar I20126 finds its massive disk hot and stable against gravitational fragmentation even as the accretion proceeds at a high rate.    
This paper is organized as fellows.  
Sect.~\ref{sec_obs} describes details of the observations and data reduction.  
We present the basic observational results in Sect.~\ref{sec_maps} and explain details of radiative transfer models for both continuum and line emissions in Sect.~\ref{sec_rtmodels}.  
We discuss the implication of our results in Sect.~\ref{sec_discussion} and conclude with a brief summary in Sect.~\ref{sec_summary}.

\section{Observations \label{sec_obs}} %
\subsection{SMA Observations} %
We carried out $345 \; \mathrm{GHz}$ observations with the SMA on 2006 July 7 and 9 in the very extended configration and 2007 August 7 in the extended configuration.  
The phase tracking center is at $(\alpha,\delta)(J2000) =$ (20:14:26.024,+41:13:32.76).  
The lower sideband covers the frequency range of $338.240 - 340.203 \; \mathrm{GHz}$ while the upper sideband covers $348.240 - 350.202 \; \mathrm{GHz}$.  
The projected baseline is in the range of $25 - 593 \; \mathrm{k}\lambda$, which is insensitive to structures larger than $3\farcs6$ (equivalent to $6 \times 10^3 \; \mathrm{AU}$).  
The system temperature varied from $320$ to $880 \; \mathrm{K}$ for the very extended configuration tracks and from $220$ to $570 \; \mathrm{K}$ for the extended configuration track.
Data inspection as well as bandpass and flux calibrations were done within the IDL superset MIR. 
The flux scale was derived from observations of the Jovian moon, Callisto, and is estimated to be accurate within 15\%.  
Temporal gains were derived from the calibrator MWC~349 and imaging were performed with the MIRIAD package.  
Both sidebands were used to generate line-free continuum maps in a multi-frequency synthesis with an effective central frequency of $344.609 \; \mathrm{GHz}$.  
The continuum image reach an rms of $4.4 \; \mathrm{mJy \, beam^{-1}}$ with a beam size of $0\farcs32 \times 0\farcs30$ and a position angle (P.A.) of $1^\circ$ using uniform weighting.  
The continuum flux density of I20126 is $1.21\; \mathrm{Jy}$.
The spectral data were gridded with a velocity resolution of $0.8 \; \mathrm{km \, s^{-1}}$ at $349.4537 \; \mathrm{GHz}$.
Imaging with the robust parameter equal to zero yields a beam size of $0\farcs37 \times 0\farcs35$ ($\mathrm{P.A.} = 4^\circ$) and an rms of $95 \; \mathrm{mJy \, beam^{-1}}$.   
The pixel size is set to be $0\farcs04$, which yields 68.0 and 90.8 pixels within one beam for the continuum and spectral line images, respectively.   

\subsection{BIMA Observations \label{sub_bima}} %
In order to find the spectral index, $\beta$, of the dust opacity in the disk, we measure the $230 \; \mathrm{GHz}$ flux density from a continuum image produced by visibilities within the same $u$-$v$ range as the SMA $345 \; \mathrm{GHz}$ observations.  
The $230 \; \mathrm{GHz}$ data were obtained with the Berkeley-Illnois-Maryland Association (BIMA) Millimeter Array on January 16 and February 14, 2002 in the A and B configuration, respectively. 
The system temperature varied from $350$ to $660 \; \mathrm{K}$ for the A configuration track and from $500$ to $840 \; \mathrm{K}$ for the B configuration track.
The calibration and imaging were performed with the MIRIAD software package.  
Temporal gains and flux density were determined by observation on the calibrator MWC~349.  
Both sidebands were used to generate the continuum map in a multi-frequency synthesis with an effective central frequency of $230.098 \; \mathrm{GHz}$.    
The continuum flux density of I20126 is $0.29 \; \mathrm{Jy}$, which is consistent with the flux density reported by \citet{Cesaroni:2014iq}. 
Comparing the flux density between $230.1$ and $344.6 \; \mathrm{GHz}$, we obtain a continuum spectral index of $3.5$, corresponding to a grain opacity spectral index of $\beta = 1.5$ in the disk.  

\footnotetext[4]{The Submillimeter Array is a joint project between the Smithsonian Astrophysical Observatory and the Academia Sinica Institute of Astronomy and Astrophysics, and is funded by the Smithsonian Institution and the Academia Sinica.}  

\section{Observational Results \label{sec_maps}} %
\begin{figure} %
\epsscale{.80} %
\plotone{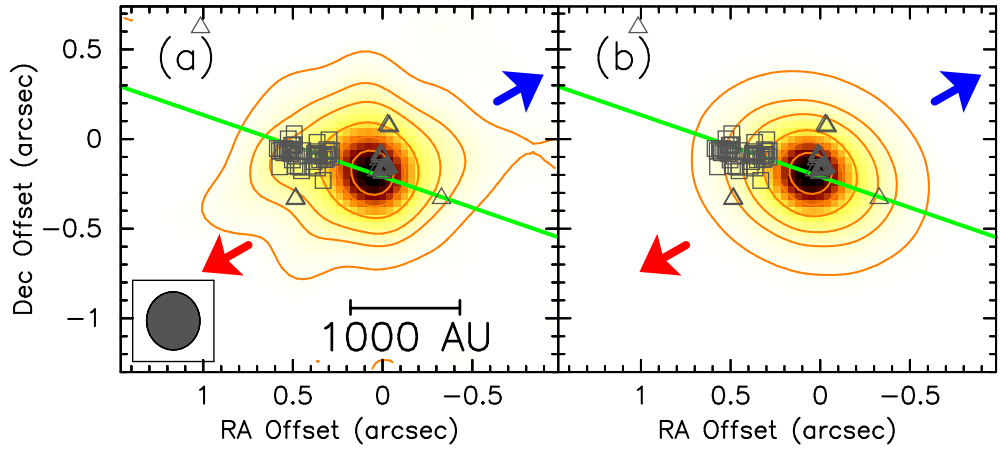} %
\caption{
$870 \; \mu\mathrm{m}$ observation and model images of dust emission.  (a) Continuum map of IRAS 20126+4104.  
The green line at P.A. of $70^\circ$ shows the disk plane in our optimized model.  
Arrows indicate the direction of the bipolar molecular outflow ($\mathrm{P.A.} = 120^\circ$).  
Contour levels are $-8.76$, 8.76, 43.8, and from 87.6 to 438 in steps of 87.6 ($20\sigma$) $\mathrm{mJy \, beam^{-1}}$ with a beam size of $0\farcs32 \times 0\farcs30$ ($\mathrm{P.A.} = 1^\circ$).  
Triangles are the positions of $\mathrm{H_2O}$ masers \citep{Edris:2005ji,Trinidad:2005ju} while squares are the positions of $\mathrm{CH_3OH}$ masers \citep{Edris:2005ji, Moscadelli:2011bo}.  
(b) Synthetic image of the optimized model for dust emission.  Contours levels follow those in (a). \label{cont}} %
\end{figure}

\begin{figure} %
\epsscale{1.0} %
\plotone{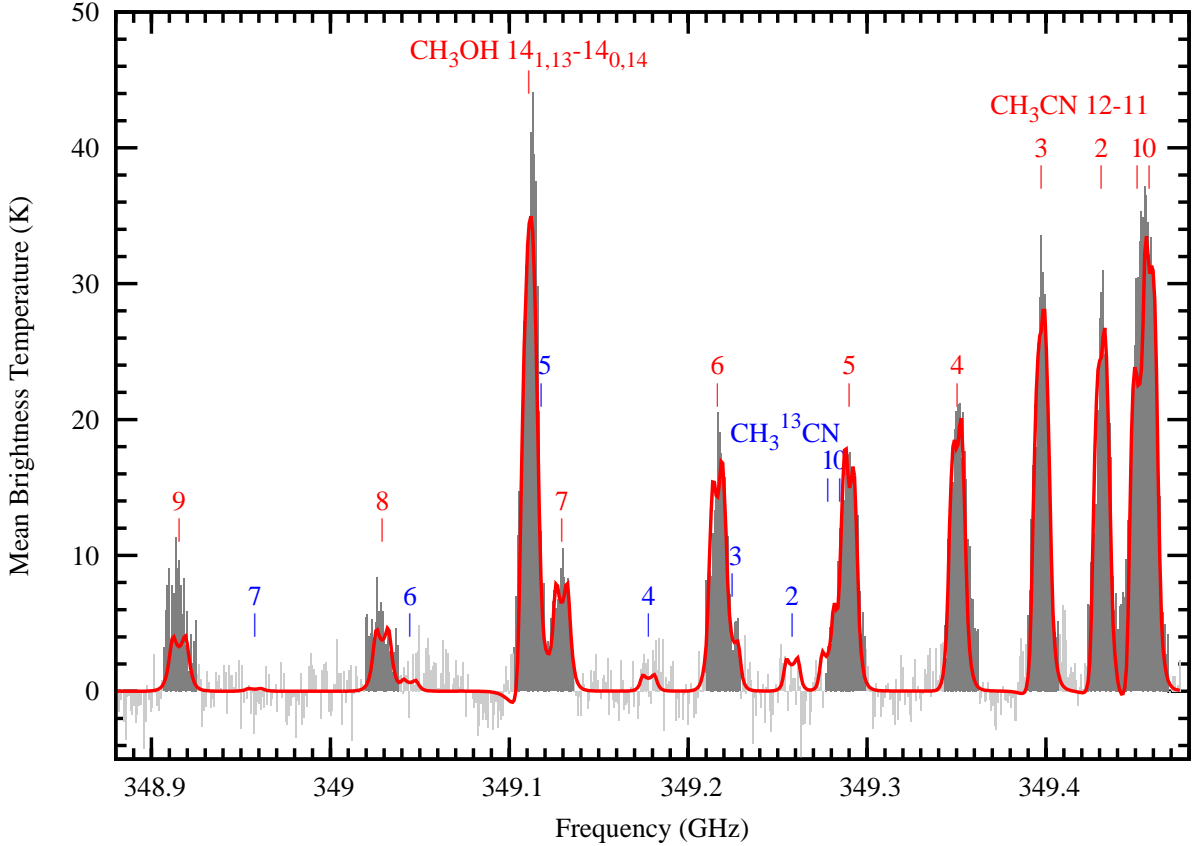} %
\caption{Mean spectrum of the observed and synthetic data obtained by averaging over the $1\farcs2$ region centered at the protostar.  The observed spectrum is shown in histogram, with channels in dark gray being considered for the model fitting and the $\chi^2$ computation.  
The red curve is the spectrum of the optimized model, which includes all the labeled 19 molecular lines.  
Undesirably, some $\mathrm{CH_3CN}$ lines are blended with lines of other species.  
For example, the $K=5$ component is blended with the $K=0,1$ components of the $\mathrm{CH_3^{13}CN}$ $J=19-18$ transition, so as the $K=6$ component with the $K=3$ component of $\mathrm{CH_3^{13}CN}$.  Meanwhile, the $K=7$ component is partially affected by the strong $\mathrm{CH_3OH} \; (14_{1,13}-14_{0,14})$ line, which also displays a strong blue-skewed spectral feature.  
\label{spec}} %
\end{figure}

\begin{figure} %
\epsscale{1.0} %
\plotone{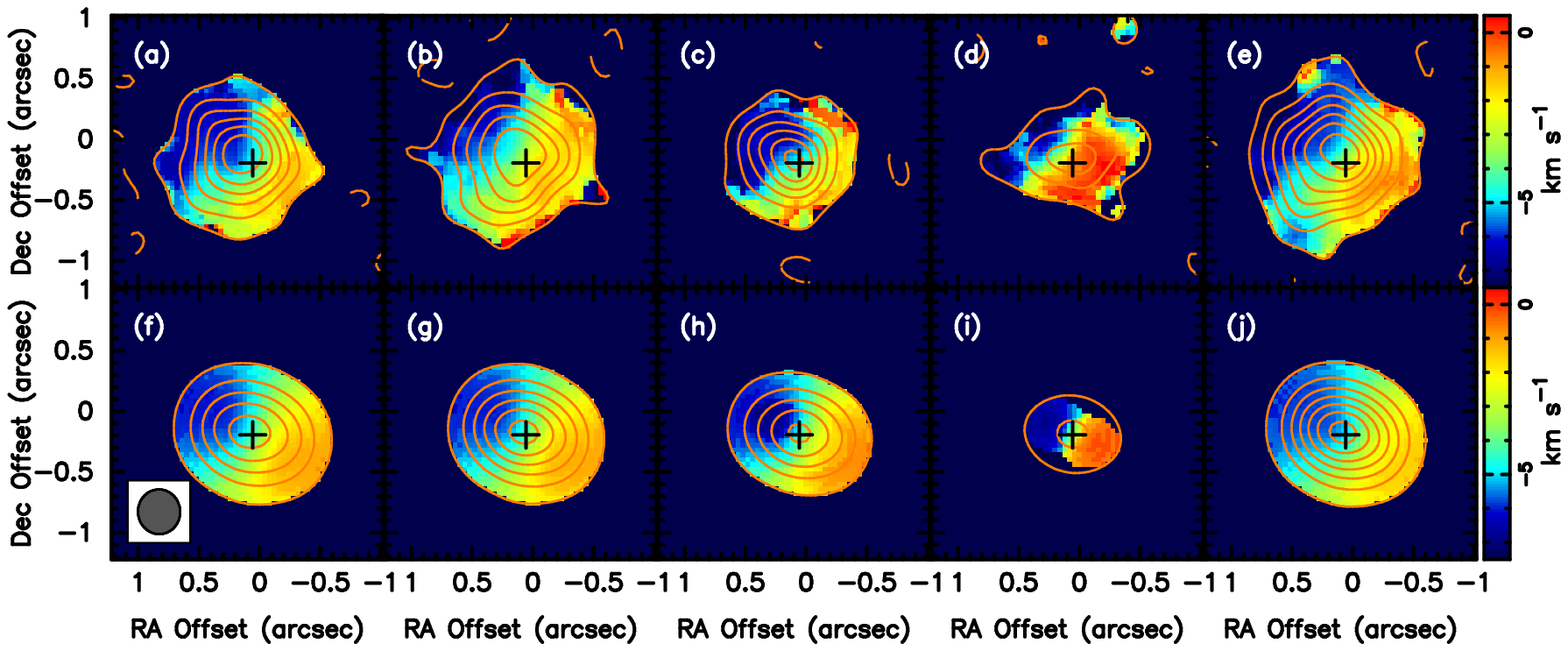} %
\caption{
Integrated intensity map (contours) overlaid on intensity-weighted velocity map (color) of observation and model images.  (a)--(d) Maps for the $K=2,3,6,9$ components of the $\mathrm{CH_3CN}$ $J=19-18$ transition.  
Contour levels are from $-0.85$ to $9.31$ in steps of $1.69 \; \mathrm{mJy \, beam^{-1} \, km \, s^{-1}}$ ($4\sigma$) with a beam size of $0\farcs37 \times 0\farcs35$ ($\mathrm{P.A.} = 4^\circ$).  
The cross marks the peak position of the continuum emission, which is assumed to be the position of the protostar.   
(e) Maps for the $\mathrm{CH_3OH}$ ($14_{1,13}-14_{0,14}$) emission.  
Contour levels are from $-0.91$ to $11.8$ in steps of $1.82 \; \mathrm{mJy \, beam^{-1} \, km \, s^{-1}}$ ($4\sigma$).
(f)--(j) Maps for the synthetic image cube of the optimized model corresponding to (a)--(e).  
 \label{ch3cnii}} %
\end{figure}

The $870 \; \mu\mathrm{m}$ dust emission (Fig.~\ref{cont}) shows a rather symmetric distribution with the peak position at $(\alpha,\delta)(J2000)=$ (20:14:26.029,+41:13:32.57), which is assumed to be the position of the protostar.  
Such symmetric morphology already suggests a disk at a moderate inclination, but the character of a rotating disk is revealed through spectral imaging.  
The observed spectrum covers the $K=0,1,\dots,9$ components of the $\mathrm{CH_3CN}$ $J=19-18$ transition with upper state energy in the range of $E_\mathrm{up}=168 - 745 \; \mathrm{K}$ and the $\mathrm{CH_3OH}$ ($14_{1,13}-14_{0,14}$) emission with $E_\mathrm{up} = 260 \; \mathrm{K}$ (Fig.~\ref{spec}).  
The $\mathrm{CH_3CN}$ lines are collisionally excited and thus sensitive to gas temperature \citep{Araya:2005gb,Chen:2006iu}.  
We can compare the Doppler shifting of spectral lines of different excitation energies to identify the spin-up of the accretion flow and the temperature gradient within disk created by the hot protostar.  
The excitation is indicated by two quantum numbers, $J$ and $K$, which specify the total angular momentum of the molecule and its projection along the principal axis.  
As examples, we showed the $K=2,3,6,9$ components of the $\mathrm{CH_3CN}$ $J=19-18$ transition (Fig.~\ref{ch3cnii}a--\ref{ch3cnii}d), whose upper states have increasing energies of $E_\mathrm{up} = 196, 232, 425, 745 \; \mathrm{K}$ and are expected to trace emission progressively close to the hot protostar.  
In addition, the $\mathrm{CH_3OH}$ ($14_{1,13}-14_{0,14}$) line shows a similar and slightly more extended emission, likely to trace part of the envelope (Fig.~\ref{ch3cnii}e).   

\begin{figure} %
\epsscale{1.0} %
\plotone{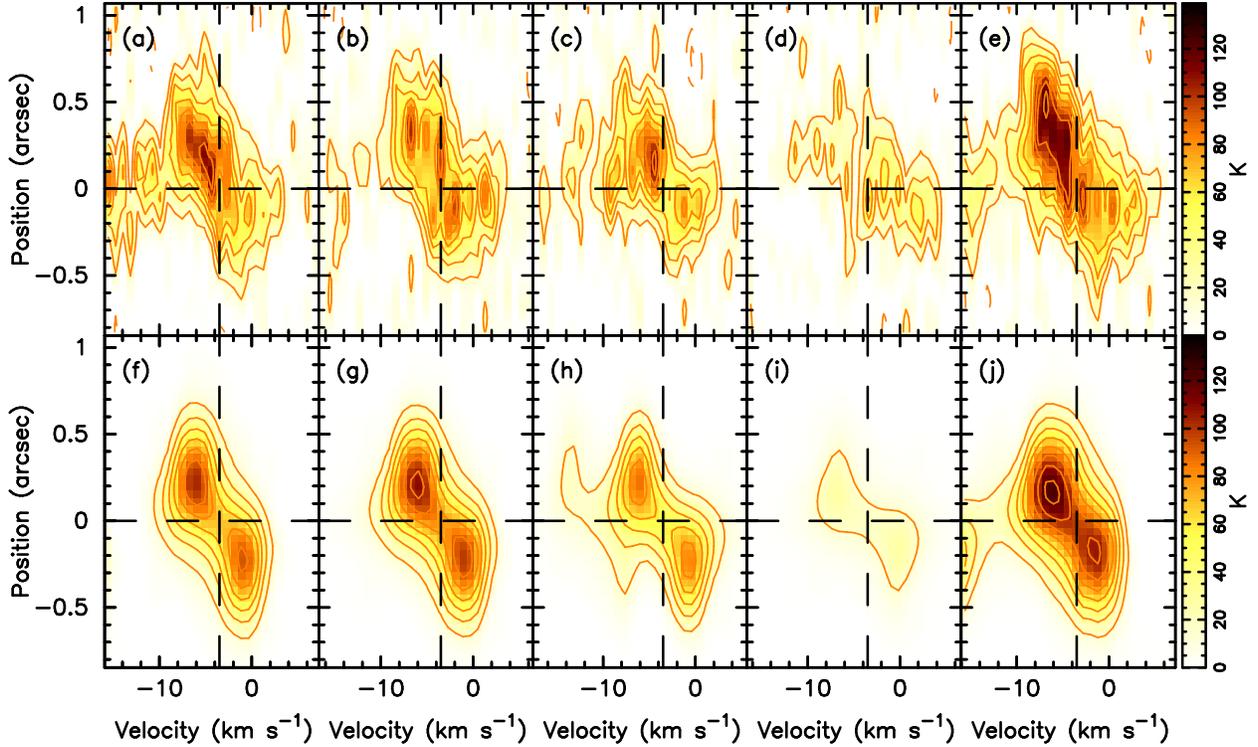} %
\caption{
Position-velocity (P-V) diagram of observation and model images.  (a)--(d) Plots for the $K=2,3,6,9$ components of the $\mathrm{CH_3CN}$ $J=19-18$ transition through the disk plane ($\mathrm{P.A.} = 70^\circ$).  
Vertical dashed line gives the systemic velocity of $-3.5 \; \mathrm{km \, s^{-1}}$ \citep{Cesaroni:1999to}. 
The position of the continuum peak is at zero position offset and indicated by horizontal dashed lines.   
Contour levels are from $-29.6$ to $103.5$ in steps of $14.8 \; \mathrm{K}$ ($2\sigma$).  
The $K=6$ component is slightly blended with the $\mathrm{CH_3^{13}CN}$ $J=19-18$ $K=3$ component appearing around $-10.6 \; \mathrm{km \, s^{-1}}$. 
(e) Plot for the $\mathrm{CH_3OH}$ ($14_{1,13}-14_{0,14}$) emission.  
Contour levels are from $-14.8$ to $118.4$ in steps of $14.8 \; \mathrm{K}$ ($2\sigma$). 
(f)--(j) Plots for the synthetic line image cube corresponding to (a)--(e).  
\label{ch3cnpv}} %
\end{figure}

The spectral line data are three-dimensional (3-D) with axes of position, position, and velocity and can be displayed in several ways.  
Plotting the average spectral line velocity (first moment) at each position shows a consistent gradient characteristic of rotation for all five observed lines (Fig.~\ref{ch3cnii}a--\ref{ch3cnii}e).  
We can select a single plane in the data cube oriented along this gradient, and plot the line intensity as a function of position and velocity (P-V diagram).  
These plots (Fig.~\ref{ch3cnpv}a--\ref{ch3cnpv}e) reveal progressively steeper velocity gradients with increasing $E_\mathrm{up}$, indicating an increase in the rotational velocity of the higher temperature gas nearer the star.  
Such spin-up motions are characteristics of an accretion flow that at least partially conserves angular momentum, the limit of which is a rotationally supported or Keplerian disk.  

The velocity gradients in the spectral line data also show the inward or radial flow of accretion in the more circular and bluer pattern of the average velocity of the $K=2$ line (Fig.~\ref{ch3cnii}a) compared with $K=9$ line (Fig.~\ref{ch3cnii}d).  
This effect arises from self-absorption of the spectral line emission in a radial flow.  
Along a line of sight through the center, a radial flow splits the line emission into red and blue shifted components from the near and far sides of the core.  
If the line were optically thin, we would see a symmetrically split profile.  
However, absorption by colder gas in the outer part of the core selectively absorbs the red-shifted emission, which is closer to its own velocity, while the emission from the far side, which is blue-shifted to a dissimilar velocity, passes through.  
If the flow were purely radial, this would produce a bullseye pattern of velocities, bluest in the center.  
In Fig.~\ref{ch3cnii}a, we see a combination of this effect along with the rotation.  
In contrast, a purely rotational flow will show a simpler one-dimensional gradient across the image as shown in Fig.~\ref{ch3cnii}d.  
This indicates that the high temperature gas has spun-up so much that the rotational velocities dominate the average.  
In I20126, the spectral line velocities and brightnesses are consistent with a more radial flow in a cooler accreting envelope that spins up and flattens to a hot rotating disk as the accretion flow approaches the star. 

\section{Radiative Transfer Models for Continuum and Line Emissions \label{sec_rtmodels}} %
The velocity patterns provide a qualitative overview of the accretion flow feeding I20126.  
We can extract precise measurements of the disk and envelope temperatures, densities, and velocities by comparing the observations with model accretion flows.  
Following earlier studies \citep{Keto:2010hj, Johnston:2011cc} that modeled lower angular resolution observations of other spectral lines, e.g. $\mathrm{NH_3}$, and the infrared continuum observations of I20126, we constructed a 3-D analytical model describing a thin accretion disk \citep{Pringle:1981gv} in Keplerian motion enveloped within the centrifugal radius of an angular-momentum-conserving accretion flow \citep{Keto:2010hj, Ulrich:1976ho}.  
We include the stellar irradiation to heat the flared disk \citep{Kenyon:1987ik} consistent with the presence of outflow cavities \citep{Qiu:2008ei, Moscadelli:2011bo}.   
A bolometric luminosity of $L_\mathrm{bol} = 1.3 \times 10^4 \; L_\odot$ is assumed \citep{Johnston:2011cc}.  
We also adopt the systemic velocity of $-3.5 \; \mathrm{km \, s^{-1}}$ \citep{Cesaroni:1999to}.

Under conditions of local thermodynamic equilibrium (LTE), we solved the radiative transfer equation for the intensity, $I_\nu$, of the continuum and spectral lines simultaneously to construct synthetic continuum images and 3-D spectral image cube for comparison with the observations.   
The radiative transfer is evaluated with the source function, i.e. the Planck function, $B_\nu(T)$, modulated by the linear sum of opacities for the continuum and spectral lines.
At the center frequency of each individual channel, we solve for $I_\nu$ with     
\begin{equation} %
    \frac{dI_\nu}{ds} = \alpha_\nu \left[ B_\nu(T) - I_\nu \right],  
\end{equation} %
where $\alpha_\nu$ is the total absorption coefficient given by 
\begin{equation} %
\alpha_\nu = \alpha_\nu^\mathrm{cont} + \alpha_\nu^\mathrm{line}.    
\end{equation} %
For the continuum emission, we have $\alpha_\nu^\mathrm{cont} = (\rho/100) \kappa_\nu$, where $\rho$ is the mass density, and a gas-to-dust mass ratio of $100$ is assumed. 
We use the dust opacity law $\kappa_\nu = 10 (\lambda/250 \mu\mathrm{m})^{-\beta} \; \mathrm{cm^2 \, g^{-1}}$ \citep{Hildebrand:1983tm} with $\beta = 1.5$ for the disk (Sec.~\ref{sub_bima}) and $\beta=1.8$ for the envelope \citep{Johnston:2011cc}.  
Regarding spectral lines, the line blending is significant for a few molecular lines so we have 
\begin{equation} %
    \alpha_\nu^\mathrm{line} = \sum_{i=1}^{N_\mathrm{line}} \alpha_{\nu}^{(i)}, 
\end{equation} %
where $N_\mathrm{line}$ is the total number of lines included in the model.  
The absorption coefficient of the $i$-th line, $\alpha_{\nu}^{(i)}$, is given by 
\begin{equation} %
\alpha_{\nu}^{(i)} = \frac{c^2}{8 \pi \nu^2} \, n X_\mathrm{mol} \, \frac{g_\mathrm{up}^{(i)} \, e^{-E_\mathrm{up}^{(i)}/k T}}{Q_\mathrm{mol}(T)} \, A_{ul}^{(i)} \left( e^{h \nu_0^{(i)}/k T} -1 \right) \Phi_\nu, 
\end{equation} %
where $n$ is the gas density, $X_\mathrm{mol}$ the abundance of the molecule, $g_\mathrm{up}^{(i)}$ the upper state degeneracy, $E_\mathrm{up}^{(i)}$ the upper state energy, $Q_\mathrm{mol}(T)$ the partition function of the molecule, $A_{ul}^{(i)}$ the spontaneous emission rate of the line, $\nu_0^{(i)}$ the rest frequency of the line, and $\Phi_\nu$ the model line profile (see Appendix~\ref{sec_lineprofile}).

The integration along line of sight is performed using the Runge-Kutta method in steps of $0.01 - 300 \; \mathrm{AU}$.  
For comparison with the observed spectral line data, synthetic spectral line images are generated through observation simulation, including visibility sampling and image making, with MIRIAD.  
The Levenberg-Marquardt method was used to optimize models by minimizing $\chi^2$ value computed with both the continuum and spectral line data.  
Only channels with emission stronger than $3\sigma$ in the central $2\arcsec$ region (dark gray histogram in Fig.~\ref{spec}) are considered for the model fitting and the $\chi^2$ computation.  
The model spectral imaging includes 19 molecular lines (Table~\ref{mod_lines}): the $K=0,1, \dots, 9$ components of the $\mathrm{CH_3CN}$ $J=19-18$ transition, the $K=0,1,\dots,7$ components of the $\mathrm{CH_3^{13}CN}$ $J=19-18$ transition, and $\mathrm{CH_3OH}$ ($14_{1,13}-14_{0,14}$).  The $\mathrm{^{12}C}/\mathrm{^{13}C}$ ratio is assumed to be 70 for a galactocentric distance of $8.3 \; \mathrm{kpc}$ \citep{Wilson:1994hu}.  
The model contains nine adjustable parameters as listed in Table~\ref{optmod}.  
The uncertainty of the parameter $a_i$ is estimated by $\Delta a_i = \pm \sqrt{C_{ii}} \sqrt{N_\mathrm{beam}} \sqrt{\Delta \chi^2}$, where $C_{ii}$ is the $i$-th diagonal term of the covariance matrix, $N_\mathrm{beam}$ is the number of pixels in the synthesized beam, and $\Delta \chi^2 \equiv \chi^2 - \chi^2_\mathrm{min} = 10.43$ gives the 68.3\% confidence level for degrees of freedom of nine.  

\begin{deluxetable}{llcccc} %
\tablewidth{0pt} %
\tablecolumns{6} %
\tablecaption{Parameters of Molecular Lines in the Models \label{mod_lines}} %
\tablehead{ \colhead{Molecule} & \colhead{Transition} & \colhead{Frequency} & \colhead{$g_\mathrm{up}$} & \colhead{$E_\mathrm{up}$} & \colhead{$\log A_{ul}$} \\
\colhead{} & \colhead{} & \colhead{(GHz)} & \colhead{} & \colhead{(K)} & \colhead{$(\log (\mathrm{s^{-1}}) )$} } %
\startdata %
$\mathrm{CH_3CN}$ & $19_9 - 18_9$ & 348.911425 & 312 & 745.41 & $-2.54355$ \\
      & $19_8 - 18_8$ & 349.024989 & 156 & 624.32 & $-2.51753$ \\
      & $19_7 - 18_7$ & 349.125301 & 156 & 517.41 & $-2.49575$ \\
      & $19_6 - 18_6$ & 349.212320 & 312 & 424.70 & $-2.47770$ \\
      & $19_5 - 18_5$ & 349.286012 & 156 & 346.22 & $-2.46297$ \\ 
      & $19_4 - 18_4$ & 349.346346 & 156 & 281.98 & $-2.45127$ \\ 
      & $19_3 - 18_3$ & 349.393298 & 312 & 232.01 & $-2.44237$ \\ 
      & $19_2 - 18_2$ & 349.426849 & 156 & 196.30 & $-2.43612$ \\ 
      & $19_1 - 18_1$ & 349.446985 & 156 & 174.88 & $-2.43241$ \\
      & $19_0 - 18_0$ & 349.453698 & 156 & 167.73 & $-2.43118$ \\
$\mathrm{CH_3^{13}CN}$ & $19_7 - 18_7$ & 348.95370 & 156 & 515.45 & $-2.49639$ \\
      & $19_6 - 18_6$ & 349.04037 & 312 & 423.24 & $-2.47834$ \\
      & $19_5 - 18_5$ & 349.11377 & 156 & 345.18 & $-2.46361$ \\
      & $19_4 - 18_4$ & 349.17386 & 156 & 281.29 & $-2.45191$ \\
      & $19_3 - 18_3$ & 349.22062 & 312 & 214.82 & $-2.44301$ \\ 
      & $19_2 - 18_2$ & 349.25403 & 156 & 179.31 & $ -2.43676$ \\ 
      & $19_1 - 18_1$ & 349.27409 & 156 & 174.76 & $-2.43305$ \\ 
      & $19_0 - 18_0$ & 349.28078 & 156 & 167.65 & $-2.43182$ \\ 
$\mathrm{CH_3OH}$ & $14_{1,13} - 14_{0,14}$ & 349.106954 & 29 & 260.20 & $-3.35597$ \\
\enddata %
\end{deluxetable} %

Analogous to the standard model of star formation with disk-mediated accretion, our model consists of a thin disk \citep{Pringle:1981gv} of mass $M_d$ in Keplerian motion around a stellar mass, $M_*$, residing within the centrifugal radius, $R_c$, of an accretion flow with constant specific angular momentum, $\Gamma$, in an infalling envelope \citep{Ulrich:1976ho}.   
To account for a fairly large disk mass, the centrifugal radius is computed with    
\begin{equation} %
R_c = \frac{\Gamma^2}{G(M_* + M_d)}.    \label{eq_Rc}  
\end{equation} %
Density-weighted means are calculated for velocity and temperature in positions where both disk and envelope are present.  
Due to the prominent outflow cavities \citep{Qiu:2008ei}, stellar irradiation is also included for anticipated heating due to the flared disk geometry \citep{Kenyon:1987ik}.  
Considering various heating processes, e.g. accretion shocks, that may occur to further raise the disk temperature, we introduce a scaling factor, $B_T$, to take these effects into account.  
Due to a fairly large disk mass, corrections for the enclosed disk mass as a function of radius are applied to rotation velocities and viscous heating produced by differential shear in the disk.

We use $r=\sqrt{x^2+y^2+z^2}$ to denote the envelope radius in spherical coordinates and $R=\sqrt{x^2+y^2}$ to denote the disk radius in cylindrical coordinates.  
Given an envelope mass accretion rate, $\dot{M}_e$, we obtain the density distribution of the envelope to be 
    \begin{eqnarray} %
    \rho_e &=& \frac{\dot{M}_e}{4\pi \sqrt{G(M_*+M_d) R_c^3}} \left( \frac{r}{R_c} \right)^{-3/2} \left( 1 + \frac{\cos \theta}{\cos \theta_0} \right)^{-1/2} \nonumber \\
    & &  \cdot \left[ 1+\left( \frac{R_c}{r} \right) \left( 3 \cos^2 \theta_0 -1 \right) \right]^{-1}, 
    \end{eqnarray} %
where $\theta_0$ is the initial polar angle of the streamline, and $r$ and $\theta$ are the polar radius and angle along a streamline.  
The envelope radius is assumed to be $0.1 \; \mathrm{pc}$, larger than the structures sensitive with our observations. 
The velocity field is described by \citep{Ulrich:1976ho,Mendoza:2004ws}
\begin{eqnarray} %
    v_r (r, \theta) &=& - v_c \left( \frac{R_c}{r} \right)^{1/2} \left( 1 + \frac{\cos \theta}{\cos \theta_0} \right)^{1/2} \nonumber \\
    v_\theta (r, \theta) &=& v_c \left( \frac{R_c}{r} \right)^{1/2} \left( \frac{\cos \theta_0 - \cos \theta}{\sin \theta} \right) \left( 1 + \frac{\cos \theta}{\cos \theta_0} \right)^{1/2} \label{venv} \\
    v_\phi(r, \theta) &=& v_c \left( \frac{R_c}{r} \right)^{1/2} \left( \frac{\sin \theta_0}{\sin \theta} \right) \left( 1 - \frac{\cos \theta}{\cos \theta_0} \right)^{1/2} \nonumber
\end{eqnarray} %
where $v_c \equiv \sqrt{G(M_*+M_d)/R_c}$ is the Keplerian velocity at $R_c$.  
In the inner region $r \lesssim R_c$, the mass of the disk gradually decreases to zero as one approaches the protostar.    
Eq.~(\ref{venv}) is an approximation for the actual velocity field that responds to the mass distribution of the disk.  
Since our observed features are mainly attributed to the disk component, we just treat the inner envelope approximately.  

The density distribution of the flared disk is described as 
    \begin{equation} %
    \rho_d (R, z) = \rho_{d0} \left( 1- \sqrt{\frac{R_*}{R}} \right) \left( \frac{R_*}{R} \right)^{2.25} e^{-z^2 / 2 H^2}, 
    \end{equation} %
where $H(R)=H_0 (R/R_*)^{1.25}$ is the disk scale-height with $H_0=0.01 R_*$, and $\rho_{d0}$ is related to the disk mass, $M_d$, by $\int_{R_\mathrm{in}}^{R_c} \int_{-\infty}^\infty \rho_d (R,z) dz \, 2\pi R dR = M_d$, where $R_\mathrm{in}$ is the inner radius of the disk.  
The surface density is defined by $\Sigma(R) \equiv \int_{-\infty}^\infty \rho_d dz \propto R^{-1}$, and the enclosed disk mass is given by $M_d(R) = \int_{R_\mathrm{in}}^R \Sigma(R^\prime) 2 \pi R^\prime dR^\prime$.  
The Keplerian rotation speed in the disk is hence 
\begin{equation} %
v_\phi(R) = \sqrt{\frac{G[M_*+M_d(R)]}{R}}. 
\end{equation} %
Given a disk mass accretion rate, $\dot{M}_d$, and the disk surface density, $\Sigma(R)$, an inward accretion velocity can be computed from     
\begin{equation} %
    v_R = -\frac{\dot{M}_d}{2 \pi R \, \Sigma(R)},   \label{eq_vR} 
\end{equation} %
which is roughly constant through the disk.  

We describe the central massive protostar as a zero-age main-sequence star \citep{Schaller:1992vq} with additional surface heating by gas accreted from the inner edge of the disk, releasing all its free-fall energy \citep{Calvet:1998fu,Johnston:2011cc}.  
The very low X-ray luminosity of $< 0.01 L_\odot$ \citep{Anderson:2011rg} suggests that the free-fall energy is largely absorbed in the stellar surface.   
Hence, the accretion luminosity that goes into heating the stellar surface is given by 
\begin{equation} %
L_\mathrm{heat} = \frac{G M_* \dot{M}_d}{R_*} \left( 1 - \frac{R_*}{R_\mathrm{in}} \right).  \label{eq_Mddot} 
\end{equation} %
The model stellar luminosity, $L_\mathrm{bol}$, including the luminosity of the protostar, $L_*$, and the accretion heating, $L_\mathrm{heat}$, is 
\begin{equation} %
L_\mathrm{bol} = L_* + L_\mathrm{heat}.   \label{eq_Lheat} 
\end{equation} %
A stellar luminosity of $L_\mathrm{bol} = 1.3 \times 10^4 \; L_\odot$ determined from infrared observations \citep{Johnston:2011cc} is applied to constrain the emerging flux at model stellar surface 
\begin{equation} %
F_\mathrm{bol} = \frac{L_\mathrm{bol}}{4 \pi R_*^2} = F_* \left( \frac{L_* + L_\mathrm{heat}}{L_*} \right),  
\end{equation} %
where $F_* \equiv L_*/4 \pi R_*^2$ and is determined by the protostellar mass, $M_*$, following relations of zero-age main-sequence stars. 

To obtain the disk temperature distribution, we consider both the accretion heating and stellar irradiation, which is responsible for the vertical thermal gradient in the disk and heating in the outer part.  
First, we compute the disk temperature at one disk scale-height, $H(R)$, with 
\begin{equation} %
    T_d(R,H) = B_T \left[ \frac{F_\mathrm{acc}(R) +F_\mathrm{irr}(R)}{\sigma_\mathrm{SB}} \right]^{1/4}, 
\end{equation} %
where $B_T$ is the temperature scaling factor, $\sigma_\mathrm{SB}$ the Stefan-Boltzmann constant, $F_\mathrm{acc}(R)$ the accretion flux \citep{Pringle:1981gv}, and $F_\mathrm{irr}(R)$ the irradiation flux \citep{Kenyon:1987ik}.   
The disk temperature at small $R$ approaches $T_d \propto R^{-3/4}$ dominated by accretion luminosity \citep{Pringle:1981gv} while the outer part is mainly heated by stellar irradiation with $T_d \propto R^{-1/2}$ \citep{Kenyon:1987ik}.  
We then obtain the vertical thermal gradient due to disk surface heating by stellar irradiation with
    \begin{eqnarray} %
    T_d(R,z) &=& T_d(R,0) \exp \left[ \ln \gamma \, \frac{|z|}{\sqrt{2} H(R)} \right]  \\
                   &=& T_d(R,H) \exp \left( -\frac{\ln \gamma}{\sqrt{2}} \right) \, \exp \left[ \ln \gamma \, \frac{|z|}{\sqrt{2} H(R)} \right]
    \end{eqnarray} %
where $\gamma$ is a parameter describing the increase in temperature with height from the mid-plane and is set to $1.5$ \citep{Dartois:2003ig}.  
The temperature profile of the envelope follows the analytical scheme \citep{Kenyon:1993fm} that gives $T_e \propto R^{-5/7}$ in the inner optically thick regime for infrared photons and $T_e \propto R^{-2/(4+\beta)}$ in the optically thin outer part with $\beta=1.8$ \citep{Johnston:2011cc}.  
The innermost dust-free zone has $T_e \propto R^{-1/2}$ and the boundary is set by dust sublimation temperature of $1600 \; \mathrm{K}$.  

The total velocity dispersion of line broadening, $\sigma$, is computed by combining the turbulent broadening of velocity dispersion, $\sigma_\mathrm{nt}$, in quadrature with the thermal broadening, $c_s \equiv \sqrt{kT/2.3 \,  m_\mathrm{H}}$, where $m_\mathrm{H}$ is the mass of Hydrogen atom.  
For the envelope, we adopted Larson’s law to describe the turbulent broadening of velocity dispersion, $\sigma_\mathrm{nt} = 1.10 (2r/1 \; \mathrm{pc})^{0.38} \; \mathrm{km \, s^{-1}}$ \citep{Larson:1981vv}. 
Since this turbulent broadening is smaller than our spectral resolution of $0.8 \; \mathrm{km \, s^{-1}}$ in most part of the envelope, we apply a line profile with instrumental broadening, $\Phi_\nu$ (see Appendix~\ref{sec_lineprofile}).     
In the disk, the Shakura-Sunyaev $\alpha$ parameter \citep{Shakura:1973uy}, is used to describe the kinematic viscosity, $\nu_k = \alpha c_s H$.  
In the thin-disk theory, the kinematic viscosity connects mass accretion rate to surface density through  
    \begin{equation} %
    \nu_k \Sigma = \frac{\dot{M}_d}{3 \pi} \left( 1-\sqrt{\frac{R_*}{R}} \right).
    \end{equation} %
Hence, we have the $\alpha$ parameter described by  
    \begin{equation} %
    \alpha = \frac{\dot{M}_d}{3 \pi \sqrt{2 \pi} \rho_{d0} c_s H_0^2} \left( \frac{R_*}{R} \right)^{1/4},  
    \end{equation} %
which just weakly depends on $T$ and $R$ with $\alpha \propto T^{-1/2} R^{-1/4}$. 
For non-thermal broadening in the disk, we found the ratio of the turbulent to thermal broadening through the $\alpha$ parameter with $\alpha = \sigma_\mathrm{nt}/c_s + B^2/4\pi \rho_d c_s^2 \approx \sigma_\mathrm{nt}/c_s$, where $B$ is the magnetic field.  
Here we assume the kinematic viscosity mainly attributed to turbulence.  
The magnetic term becomes important if the field strength is greater than a critical value of $B_\mathrm{crit} = \sqrt{4\pi \rho_d c_s^2} \geq 5 \; \mathrm{mG}$ at $R_c$.  
Given the typical field strength of $\leq 1 \; \mathrm{mG}$ in star-forming cores \citep{Crutcher:2010ir}, the approximation of the $\alpha$ parameter for deriving the turbulent broadening is reasonable in I20126.  
We kept $\alpha \leq 1$ in the disk to make turbulence subsonic when optimizing models.  

\section{Results and Discussion \label{sec_discussion}} %
The model is specified by nine adjustable parameters, listed in Table~\ref{optmod}, that are optimized by $\chi^2$ minimization.  
Since our observations are insensitive to $R_\mathrm{in}$, we do not intend to determine its value but assume $R_\mathrm{in} = 5 R_*$ based on previous infrared studies \citep{Johnston:2011cc} and findings in disks around low-mass stars \citep{Shu:1994gr}.     
The best-fit model gives a disk mass of $1.5 \, M_\odot$ and a centrifugal radius, $R_c$, of $858 \; \mathrm{AU}$ rotating about a $12.0 \; M_\odot$ protostar with an accretion luminosity of $2.7 \times 10^3 \; L_\odot$, giving a disk mass accretion rate, $\dot{M}_d$, of $3.9 \times 10^{-5} \; M_\odot \, \mathrm{yr^{-1}}$.  
The optimization obtains a reduced $\bar{\chi}^2$ value of 1.6, which is optimized with $661696$ pixels including both the continuum and spectral line data, equivalent to $7300$ independent data points.  
The derived parameters along with their uncertainties are listed in Table~\ref{modpar}.  
The synthetic continuum image from the best-fit model is shown in Fig.~\ref{cont}b, while the integrated intensity images  and the P-V diagrams of the $\mathrm{CH_3CN}$ and $\mathrm{CH_3OH}$ emissions are shown in Fig.~\ref{ch3cnii}f--\ref{ch3cnii}j and Fig.~\ref{ch3cnpv}f--\ref{ch3cnpv}j, respectively.    
For choices of $R_\mathrm{in}$ in the range of $2R_* - 10 R_*$, parameters of the best-fit models vary slightly within 4\% of those listed in Table~\ref{optmod} with the exception of $B_T$, which compensates the change of $R_\mathrm{in}$ and varies within 30\%. 

\begin{deluxetable}{cc} %
\tablewidth{0pt} %
\tablecolumns{2} %
\tablecaption{Parameters of the Optimized Model  \label{optmod}} %
\tablehead{ \colhead{Parameter} & \colhead{Value} } %
\startdata %
Stellar mass, $M_*$  & $12.0 \pm 0.2 \; M_\odot$ \\
Disk mass, $M_d$ & $1.5 \pm 0.1 \; M_\odot$ \\
Inclination angle\tablenotemark{\dagger}, $i$  & $48^\circ \pm 2^\circ$ \\
Position angle & $70^\circ \pm 2^\circ$ \\
Disk temperature scaling factor, $B_T$ & $1.5 \pm 0.1$ \\
Specific angular momentum, $\Gamma$ & $(3.20 \pm 0.04) \times 10^3 \; \mathrm{AU \, km \, s^{-1}}$ \\
Envelope accretion rate, $\dot{M}_e$ & $(1.5 \pm 0.2) \times 10^{-3} \; M_\odot \, \mathrm{yr^{-1}}$ \\
$\mathrm{CH_3CN}$ fractional abundance, $X_\mathrm{CH_3CN}$ & $(2.4 \pm 0.3) \times 10^{-8}$ \\
$\mathrm{CH_3OH}$ fractional abundance, $X_\mathrm{CH_3OH}$ & $(1.5 \pm 0.3) \times 10^{-6}$ 
\enddata %
\tablenotetext{\dagger}{Inclination angle is defined as the angle between the disk axis and the line of sight.} %
\end{deluxetable} %

The accretion timescale is estimated by 
\begin{equation} %
\tau_\mathrm{acc} = \frac{R}{v_R},  \label{eq_tauacc}
\end{equation} %
which gives the time for a mass element in the disk to be accreted, and $\tau_\mathrm{acc} \propto R$.  
The rotation period of a Keplerian disk is 
\begin{equation} %
P_\mathrm{rot} = 2\pi \sqrt{ \frac{R^3}{G[M_*+M_d(R)]} } \quad \propto R^{3/2}.  \label{eq_Prot}
\end{equation} %
Hence the ratio $\tau_\mathrm{acc}/P_\mathrm{rot} \propto R^{-1/2}$ and reaches a minimum at $R_c$.  
As long as $\tau_\mathrm{acc} > P_\mathrm{rot}$ at $R_c$, the accretion timescale is longer than the rotation period at all radii, and the accretion is mediated by a rotationally supported disk.  
In I20126, the surface density at $R_c$ is $2.8 \; \mathrm{g \, cm^{-2}}$, which implies an inward velocity, $v_R$, of $0.11 \; \mathrm{km \, s^{-1}}$.  
The accretion timescale at $R_c$, $\tau_\mathrm{acc} = R_c/v_R$, is $3.7 \times 10^4 \; \mathrm{yr}$, longer than the rotation period of $6.9 \times 10^3 \; \mathrm{yr}$, so the accretion is mediated by a rotationally supported disk.  

\begin{figure} %
\epsscale{1.0} %
\plotone{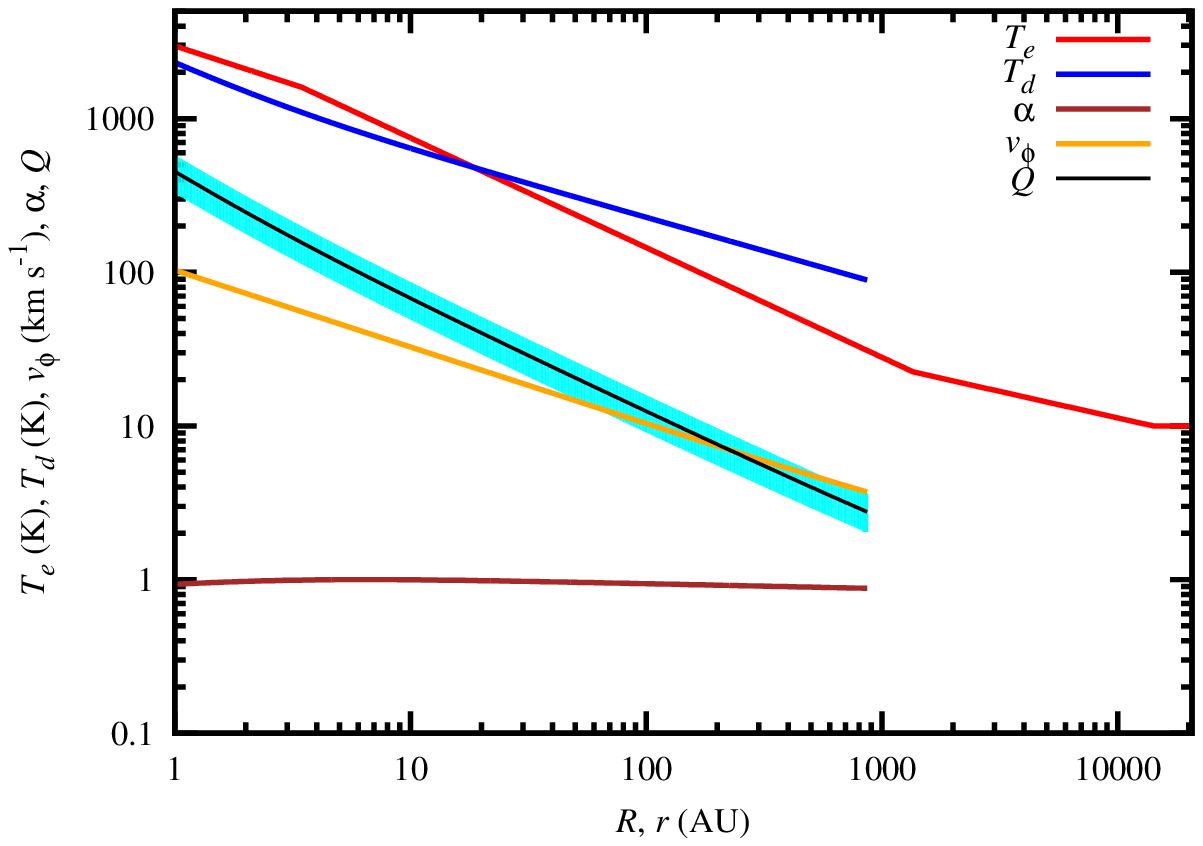} %
\caption{Parameters in the optimized model versus radius, including the envelope temperature, $T_e$ (red), the disk temperature $T_d$ (blue), the rotational velocity in the disk, $v_\phi$ (orange), the Shakura-Sunyaev $\alpha$ parameter (brown), and the Toomre-$Q$ (black) with uncertainty range (cyan).  
Values in the mid-plane of the disk parameters are plotted.  
The Toomre-$Q$ is everywhere larger than $2.8$, which makes the disk stable to gravitational instability.  
The fractional uncertainty of Toomre-$Q$ is about 27\% through the disk. 
Disk turbulence is assumed to be subsonic, which constrains $\alpha \le 1$ in model optimization.  
The disk mid-plane temperature profile takes an asymptotic form $T_d \propto R^{-3/4}$ for small R where the accretion luminosity dominates and $T_d \propto R^{-1/2}$ for large R where stellar irradiation is important.  
For the envelope temperature distribution, the inner region is optically thick for infrared photons with $T_e \propto r^{-5/7}$ while the outer optically thin region has $T_e \propto r^{-0.34}$.  
The minimum temperature is set to be $10 \; \mathrm{K}$ in the outermost part of the envelope.  
\label{para}} %
\end{figure}

As the stellar luminosity is actually a proxy for the stellar mass, our model derives a larger stellar mass, leading to a  moderately inclined disk at an angle of $i = 48^\circ$ rather than an edge-on geometry, i.e. $i=90^\circ$ \citep{Cesaroni:2005eu,Cesaroni:2014iq}.  
A smaller inclination angle of $41^\circ$ has also been suggested by numerical simulations with a misaligned magnetic field with respect to the disk rotation axis \citep{Shinnaga:2012kh}.  
The temperature of the disk is above $90 \; \mathrm{K}$ at all radii and becomes warmer than the enveloping accretion flow beyond $19 \; \mathrm{AU}$.  
Figure~\ref{para} shows the temperature distributions of the disk and the envelope.  
The mid-plane density of the disk is everywhere larger than $1.6 \times 10^8 \; \mathrm{cm^{-3}}$ and is significantly higher than the critical density for collisional de-excitation for thermalization of the $\mathrm{CH_3CN}$ lines, which is about $5 \times 10^6 \; \mathrm{cm^{-3}}$ at $100 \; \mathrm{K}$.  
The densities in the envelope are lower and thus may not fully satisfy the LTE conditions.  
Since the observed lines are dominated by the disk component, the approximate treatment of the envelope is not significant. 

The derived gas density and mass depend on the molecular abundances, which are assumed to be constant in the entire system, including the disk and envelope.  
While variations of the molecular abundance cannot be ruled out, the fact that the derived temperature exceeds ice mantle sublimation point readily implies the enhancement of $\mathrm{CH_3CN}$ in the gas phase.  
The derived $\mathrm{CH_3CN}$ abundance of $2.4 \times 10^{-8}$ is comparable with those found in hot molecular cores \citep{HernandezHernandez:2014pi}.  
An abrupt jump of $\mathrm{CH_3CN}$ abundance within the domain of interests is therefore not expected. 

\begin{deluxetable}{ccc} %
\tablewidth{0pt} %
\tablecolumns{3} %
\tablecaption{Parameters Derived from the Optimized Model \label{modpar}} %
\tablehead{ \colhead{Parameter} & \colhead{Value} & \colhead{Equation} } %
\startdata %
Centrifugal radius, $R_c$ & $858 \pm 26 \; \mathrm{AU}$  & (\ref{eq_Rc}) \\  
Accretion luminosity, $L_\mathrm{heat}$ & $(2.7 \pm 0.5) \times 10^3 \; L_\odot$ & (\ref{eq_Lheat}) \\
Disk accretion rate, $\dot{M}_d$ & $(3.9 \pm 0.8) \times 10^{-5} \; M_\odot \, \mathrm{yr^{-1}}$ &  (\ref{eq_Mddot}) \\
Inward accretion velocity, $v_R$ & $0.11 \pm 0.03 \; \mathrm{km \, s^{-1}}$ & (\ref{eq_vR}) \\
\hline
\sidehead{Representative value at $R_c$} %
\hline
Surface density, $\Sigma$ & $2.8 \pm 0.3 \; \mathrm{g \, cm^{-2}}$ & \\
Accretion timescale, $\tau_\mathrm{acc}$ & $(3.7 \pm 0.9) \times 10^4 \; \mathrm{yr}$ & (\ref{eq_tauacc}) \\
Rotational period, $P_\mathrm{rot}$ & $(6.9 \pm 0.07) \times 10^3 \; \mathrm{yr}$ & (\ref{eq_Prot}) \\
Toomre-$Q$ parameter & $2.8 \pm 0.8$ & (\ref{eq_Q}) 
\enddata %
\end{deluxetable} %

Having determined the properties of the disk by comparison with our observations, we can assess its dynamical stability.  
We calculate the Toomre-$Q$ parameter, $Q = c_s \Omega/\pi G \Sigma$, and find it larger than $2.8$ everywhere in the disk, which makes the disk stable to fragmentation (Fig.~\ref{para}).  
The fractional uncertainty of Toomre-$Q$ is about 27\% through the disk. 
The Shakura-Sunyaev $\alpha$ parameter \citep{Shakura:1973uy} is $\sim 1$ within $10 \; \mathrm{AU}$ and decreases slowly to $0.88$ at $R_c$.  
The value of $\alpha$ is much higher than the typical value of $\alpha = 0.1$, above which fragmentation occurs in simulations of isolated disks with simple cooling laws \citep{Rice:2005fl}.  
Yet numerical models appropriate for high-mass protostars tend to find higher $\alpha$ \citep{Kratter:2010gs}.  
A direct comparison between observations and simulations may still be premature because the evolution of the accretion flow in I20126 may not necessarily follow the evolution assumed for the simulations \citep{Kratter:2008bs}. 

\section{Summary \label{sec_summary}} %
We present new $345 \; \mathrm{GHz}$ continuum and spectral line observations of the disk around the high-mass protostar I20126 with the SMA that achieved the highest angular resolution ($0\farcs37$, equivalent to $\sim 600 \; \mathrm{AU}$) to resolve the accretion flow. 
The continuum emission shows a fairly symmetric morphology, which suggests the disk at a moderate inclination angle.
Observations of $\mathrm{CH_3CN}$ and $\mathrm{CH_3OH}$ lines resolve the central $2000 \; \mathrm{AU}$ region, where the kinematics display a clear rotation velocity pattern.   
Position-velocity diagrams of the $\mathrm{CH_3CN}$ lines reveal progressively steeper velocity gradients with increasing upper state energy, indicating an increase in the rotational velocity of the hotter gas nearer the protostar.  
Such spin-up motions are characteristics of a rotationally supported disk.  

We assess the dynamical stability of this massive disk through the Toomre-$Q$ parameter.  
To evaluate the Toomre-$Q$ as a function of radius through the disk, we measure the gas density, temperature, and rotational velocity in the disk by comparing data with synthetic data generated by radiative transfer models analogous to the standard model of star formation with disk-mediated accretion.   
Given a luminosity of $1.3 \times 10^4 \; L_\odot$, the optimized model finds a disk mass of $1.5 \; M_\odot$ and a centrifugal radius of $858 \; \mathrm{AU}$ rotating about a $12.0 \; M_\odot$ protostar with a disk mass accretion rate of $3.9 \times 10^{-5} \; M_\odot \, \mathrm{yr^{-1}}$.  
These physical conditions render $Q > 2.8$ everywhere in the disk, which makes the disk stable to fragmentation.  
 
Our high angular resolution SMA observations of I20126 provide evidence for a stable massive accretion disk around a high-mass protostar.  
In contrast to some theoretical expectations of massive disks prone to local instabilities, the disk of I20126 is found to be hot and stable to fragmentation even as the accretion proceeds at a high rate.  
Such conditions may help to maintain the disk around massive stars and preserve opportunities for developing companions or a planetary system in a later phase of the protostellar evolution.  

\acknowledgments %
This work is supported by the Taiwan Ministry of Science and Technology, project MOST 103-2119-M-007-006-MY3.  


\appendix %
\section{Instrumental Broadening Using a Boxcar Function \label{sec_lineprofile}} %
\begin{figure} %
\epsscale{1.0} %
\plotone{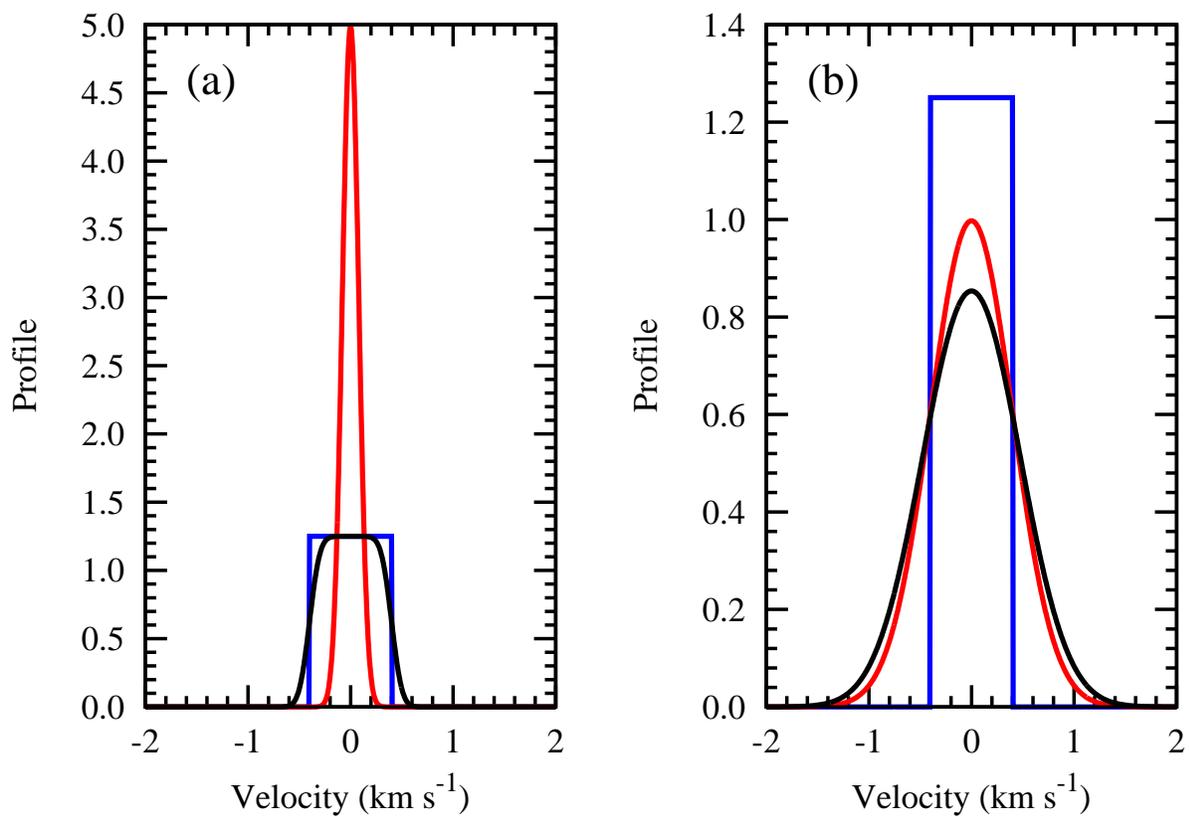} %
\caption{
(a) Model profile function, $\Phi(v)$ (black), given by the convolution of a boxcar function, ${\cal P}(v)$ (blue), with a channel spectral resolution of $\Delta = 0.8 \; \mathrm{km \, s^{-1}}$ and a gaussian function, $\phi(v)$ (red), with a relatively small line broadening of $\sigma = 0.08 \; \mathrm{km \, s^{-1}}$.  (b) Similar plot but for a gaussian function with a comparable line broadening of $\sigma = 0.4 \; \mathrm{km \, s^{-1}}$.
\label{boxgau}} %
\end{figure}

Since the turbulence line width in the inner part of the envelope is smaller than the spectral resolution of our observations, it is necessary to account for the instrumental broadening in our models.  
The model line profile, $\Phi(v)$, is given by the convolution of a boxcar, ${\cal P}(u)$, set by channel spectral resolution, $\Delta$, and a gaussian function, $\phi(v)$, of line broadening, $\sigma$, set by thermal broadening, $c_s$, and nonthermal broadening, $\sigma_\mathrm{nt}$, with $\sigma^2 = c_s^2 + \sigma_\mathrm{nt}^2$.  
The intrinsic gaussian line profile is given by 
\begin{equation} %
\phi(v) = \frac{1}{\sqrt{2\pi} \sigma} \exp \left[ {-\frac{(v - v_0)^2}{2 \sigma^2}} \right] 
\end{equation} %
where $v_0$ is the systemic velocity.  
The frequency response of one channel is approximated by a boxcar function 
\begin{equation} %
{\cal P}(u) = \left\{ \begin{array}{ll}  1/\Delta & \quad \quad \mathrm{for} -\Delta/2 \leqslant u \leqslant \Delta/2, \\
                                               0 & \quad \quad \mathrm{otherwise.} \end{array} \right.
\end{equation} %
Hence, we calculate the line profile with the instrumental broadening by convolving $\phi(v)$ with ${\cal P}(u)$ 
\begin{eqnarray*} %
\Phi(v) \equiv {\cal P} \otimes \phi(v) &=& \int_{-\frac{\Delta}{2}}^{\frac{\Delta}{2}}  \frac{1}{\sqrt{2\pi} \sigma \Delta} \exp \left[ {-\frac{(v - u - v_0)^2}{2 \sigma^2}} \right] du \\
 &=& \frac{1}{2\Delta} \left[ \mathrm{erf}\left( \frac{\Delta/2-(v-v_0)}{\sqrt{2} \sigma} \right) - \mathrm{erf}\left( \frac{-\Delta/2-(v-v_0)}{\sqrt{2} \sigma} \right) \right], 
\end{eqnarray*} %
where $\mathrm{erf}(x)$ is the error function.  
Since a simple formula is available for the complementary error function, $\mathrm{erfc}(x) \equiv 1 - \mathrm{erf}(x)$, one can also rewrite the line profile as 
\begin{equation} %
\Phi(v) = \frac{1}{2 \Delta} \left[ \mathrm{erfc}\left( \frac{-\Delta/2 - (v - v_0)}{\sqrt{2} \sigma} \right) - \mathrm{erfc}\left( \frac{\Delta/2 - (v - v_0)}{\sqrt{2} \sigma} \right) \right]. 
\end{equation} %
Examples of $\Phi(v)$ are shown in Fig.~\ref{boxgau} for the case of a spectral resolution of $\Delta = 0.8 \; \mathrm{km \, s^{-1}}$, same as our observations.  

\bibliographystyle{apj} %
\bibliography{myrefs} %




\end{document}